\documentclass[a4paper,12pt]{article}

\usepackage[dvips]{graphicx}
\usepackage{geometry}
\usepackage[centertags]{amsmath}
\usepackage{amsfonts}
\usepackage{amssymb}
\usepackage{amsthm}
\usepackage{latexsym}
\usepackage{psfrag}

\newcommand{\uD}{\ensuremath{\mathrm{d}}}
\newcommand{\uE}{\ensuremath{\mathrm{e}}}
\newcommand{\uI}{\ensuremath{\mathrm{i}}}

\begin{document}

%\sloppy

\title{\textbf{Explicit expression for the photon number emission in
synchrotron radiation}\thanks{Published in \textit{Physics~Letters}~\textbf{A268}
(2000) 35--36.}}

\author{\textsc{E.~B.~Manoukian}\thanks{E-mail: \texttt{edouard@ccs.sut.ac.th}} \ and
\ \textsc{N. Jearnkulprasert} \\
{School of Physics, \ Suranaree University of Technology} \\
\ Nakhon Ratchasima, 30000, Thailand }
\date{} \maketitle

\begin{abstract}
An explicit and remarkably simple one-dimensional integral
expression is derived for the mean number $\left<N\right>$ of
photons emitted per revolution in synchrotron radiation\@.  The
familiar high-energy expression $5\pi\alpha/\sqrt{3(1-\beta^2)}$,
printed repeatedly in the literature, is found to be inaccurate
and only truly asymptotic with relative errors of $160\%$, $82\%$
for $\beta=0.8$, $0.9$, respectively\@. A new improved high-energy
expression for $\left<N\right>$ is given\@. \\
\end{abstract}

The fascinating story of synchrotron radiation emphasizing both
the early theoretical and experimental developments is well
documented in the literature (e.g., \cite{B01})\@.  The monumental
pioneering theoretical contribution of Schwinger \cite{B02,B03}
and its direct experimental impact has been particularly noted
\cite{B01}\@.  Although the main features of synchrotron radiation
have been well known for a long time, there is certainly room for
further developments and improvements\@.   In this Letter, we use
an earlier integral for the power of radiation obtained by
Schwinger \cite{B02,B03,B05} over fifty years ago to derive an
explicit expression for the mean number $\left<N\right>$ of
photons emitted per revolution with no approximation made\@.  The
derived result for $\left<N\right>$ is a remarkably simple
one-dimensional integral\@.  We infer that the familiar
high-energy expression $5\pi\alpha/\sqrt{3(1-\beta^2)}$
\cite{B06,B07} for $\left<N\right>$, and repeatedly printed in the
literature, is rather inaccurate and is to be considered only as
truly asymptotic in the sense that even for speeds $\beta=0.9$,
$0.8$ deviations from this expression are rather significant with
large relative errors of $82\%$, $160\%$ (!), respectively\@.  In
particular, our explicit result for $\left<N\right>$ is used to
obtain a much-improved asymptotic high-energy expression for
radiating particles\@. \\

Our starting expression for $\left<N\right>$ is obtained directly
from Schwinger's formulae (\cite{B02}, Eqs.~III~6, 7; \cite{B04},
Eq.~(C.11)) for the power\@.
\begin{equation}\label{Eqn01}
  \left<N\right>=\dfrac{\alpha}{\beta}\int^{\infty}_{0}\!\!\uD{}z
  \int^{\infty}_{-\infty}\!\!\uD{}x\;\uE^{-\uI{}zx}(\beta^{2}\cos{}x-1)
  \dfrac{\sin\left(2\beta{}z\sin\dfrac{x}{2}\right)}{\sin\dfrac{x}{2}}.
\end{equation}
Since the integrand factor in (\ref{Eqn01}) multiplying
$\exp(-\uI{}zx)$ is an \underline{even} function of $x$, only the
real part of the integral is non-vanishing\@.  It is easily
verified that $\left<N\right>=0$ for $\beta=0$, as it should be,
when integrating over $x$ and $z$ in (\ref{Eqn01}) and using in
the process that $\int^{\infty}_{0}\!\!\uD{}z\;z
\int^{\infty}_{-\infty}\!\!\uD{}x\;\uE^{-\uI{}zx}$\@.  Accordingly
we may rewrite (\ref{Eqn01}) as
\begin{align}\label{Eqn02}
  \left<N\right> &= \alpha\int^{\infty}_{0}\!\!\uD{}z\int^{\infty}_{-\infty}\!\!\uD{}x\;
  \uE^{-\uI{}zx}\int^{\beta}_{0}\!\!\uD\rho \nonumber \\
  &\qquad\times
  \left(\cos{}x\dfrac{\sin\left(2\beta{}z\sin\dfrac{x}{2}\right)}{\sin\dfrac{x}{2}}
  -\dfrac{2z}{\beta}\left[\cos\left(2z\rho\sin\dfrac{x}{2}\right)-1\right]\right)
\end{align}
conveniently written by taking into account the explicit vanishing
property of $\left<N\right>=0$ for $\beta=0$\@.    To evaluate
$\left<N\right>=0$, we first integrate over $z$, then over $\rho$
to obtain
\begin{equation}\label{Eqn03}
  \left<N\right>=\alpha\int^{\infty}_{-\infty}\!\!\uD{}x
  \left\{\dfrac{2(1-\beta^2\cos{}x)}{\left(x^2-4\beta^2\sin^2\dfrac{x}{2}\right)}
  -\dfrac{2}{x^2}\right\}.
\end{equation}
Upon a change of variable $x/2\to{}x$, we finally obtain the
remarkably simple expression
\begin{align}\label{Eqn04}
  \left<N\right> &= 2\alpha\beta^2\int^{\infty}_{0}\!\dfrac{\uD{}x}{x^2}
  \left[\dfrac{\left(\dfrac{\sin{}x}{x}\right)^{2}-\cos(2x)}{\left(1-\beta^2
  \left(\dfrac{\sin{}x}{x}\right)^{2}\right)}\right] \nonumber \\
  &\equiv \alpha{}f(\beta).
\end{align}
There is no question of the existence of the latter integral for
all $0\leq\beta<1$\@.  [For completeness we provide values for
$f(\beta)$: $0.1731$, $0.7694$, $2.1351$, $5.7951$, $11.4003$,
$54.7651$ corresponding, respectively, to $\beta=0.2$, $0.4$,
$0.6$, $0.8$, $0.9$, $0.99$.] \\

Eq.~(\ref{Eqn04}) leads to the following expression~:
\begin{align}\label{Eqn0567}
  f(\beta) &= f_{0}(\beta)+a_0+\varepsilon(\beta) \\
  f_{0}(\beta) &= 10\beta\int^{\infty}_{0}\!\!\uD{}x
  \left[3(1-\beta^2)+\beta^2x^2\right]^{-1} \nonumber \\
  &= \frac{5\pi}{\sqrt{3(1-\beta^2)}}. \\
  a_{0} &= 2\int^{\infty}_{0}\!\dfrac{\uD{}x}{x^2}
  \dfrac{\left[6\left(\dfrac{\sin{}x}{x}\right)^{2}-\cos(2x)-5\right]}{\left[1-
  \left(\dfrac{\sin{}x}{x}\right)^{2}\right]} \nonumber \\
  &= {-9.5580}.
\end{align}
For $\beta\to{}1$
\begin{equation}\label{Eqn08}
  \varepsilon(\beta)=\mathcal{O}\left(\sqrt{1-\beta^2}\right).
\end{equation}
That is, at high energies we may write
\begin{equation}\label{Eqn09}
  \left<N\right> \cong \dfrac{5\pi\alpha}{\sqrt{3(1-\beta^2)}}+a_{0}\alpha.
\end{equation}
The asymptotic constant $a_{0}$ is overwhelmingly large in
magnitude\@. It is the important contribution that survives in the
limit $\beta\to{}1$ beyond the $1/\sqrt{1-\beta^2}$ term\@.
Eq.~(\ref{Eqn09}) provides a significantly much improved
high-energy expression for $\left<N\right>$\@. The relative errors
in (\ref{Eqn09}) are quite satisfactory with $4.11\%$, $1.34\%$,
$0.063\%$ for $\beta=0.8$, $0.9$, $0.99$\@.  They are to be
compared with the relative errors of $160\%$, $82\%$, $17\%$,
respectively, for the truly asymptotic earlier formula\@. \\

\end{document}